\documentclass[aps,prd,twocolumn,nofootinbib,superscriptaddress]{revtex4-2}

\usepackage{amsfonts}
\usepackage{mathrsfs}
\usepackage{graphicx}
\usepackage{amsmath}
\usepackage{amssymb}
\usepackage{color}

\allowdisplaybreaks[4]

\usepackage[colorlinks=true,linkcolor=blue,citecolor=blue, urlcolor=blue]{hyperref}

\begin{document}

\title{Improved analysis of the decay width of $t\to Wb$ up to N$^{3}$LO QCD corrections}

\author{Jiang Yan}
\email{yjiang@cqu.edu.cn}
\address{Department of Physics, Chongqing Key Laboratory for Strongly Coupled Physics, Chongqing University, Chongqing 401331, P.R. China}

\author{Xing-Gang Wu}
\email{wuxg@cqu.edu.cn}
\address{Department of Physics, Chongqing Key Laboratory for Strongly Coupled Physics, Chongqing University, Chongqing 401331, P.R. China}

\author{Hua Zhou}
\email{zhouhua@swust.edu.cn}
\address{School of Science, Southwest University of Science and Technology, Mianyang 621010, P.R. China}

\author{Hong-Tai Li}
\email{liht@cqu.edu.cn}
\address{Department of Physics, Chongqing Key Laboratory for Strongly Coupled Physics, Chongqing University, Chongqing 401331, P.R. China}

\author{Jing-Hao Shan}
\email{sjh@cqu.edu.cn}
\address{Department of Physics, Chongqing Key Laboratory for Strongly Coupled Physics, Chongqing University, Chongqing 401331, P.R. China}

\date{\today}

\begin{abstract}

In this paper, we analyze the top-quark decay $t\to Wb$ up to next-to-next-to-next-to-leading order (N$^{3}$LO) QCD corrections. For the purpose, we first adopt the principle of maximum conformality (PMC) to deal with the initial pQCD series. Then we adopt the Bayesian analysis approach, which quantifies the unknown higher-order terms' contributions in terms of a probability distribution, to estimate the possible magnitude of the uncalculated N$^{4}$LO-terms. In our calculation, an effective strong coupling constant $\alpha_{s}(Q_{*})$ is determined by using all non-conformal $\{\beta_{i}\}$ terms associated with the renormalization group equation. This leads to a next-to-leading-log PMC scale $Q_{*}^{(\rm NLL)}=10.3048$ GeV, which can be regarded as the correct momentum flow of the process. Consequently, we obtain an improved scale-invariant pQCD prediction for the top-quark decay width, e.g. $\Gamma_{t}^{\rm tot} = 1.3120 \pm 0.0038$ GeV, whose error is the squared average of the uncertainties from the decay width of $W$-boson $\Delta \Gamma_{W} = \pm 0.042$ GeV, the coupling constant $\Delta \alpha_{s}(m_{Z}) = \pm 0.0009$, and the predicted N$^{4}$LO-terms. The magnitude of the top-quark pole mass greatly affects the total decay width. By further taking the PDG top-quark pole mass error from cross-section measurements into consideration, e.g. $\Delta m_{t} = \pm 0.7$ GeV, we obtain $\Gamma_{t}^{\rm tot} = 1.3120 ^{+0.0194}_{-0.0192}$ GeV.

\end{abstract}

\maketitle

\section{Introduction}

The top quark is the heaviest elementary particle in the Standard Model~(SM), and it is remarkable for its decay processes. Compared to other quarks, the top quark has a much larger mass and a significantly shorter lifetime. It does not have enough time to form any hadron before decaying itself. The top quark's substantial Yukawa coupling with the Higgs boson exerts considerable influence on the SM observables. Furthermore, it serves as an exceptional laboratory for probing fundamental interactions at the Electroweak~(EW) symmetry-breaking scale and beyond.

Within the SM, the top quark decays almost exclusively into a $W$-boson and a $b$-quark. Thus the top-quark total decay width can be deduced from the partial decay width $\Gamma(t\to Wb)$ and the branching fraction $\mathcal{B}(t\to Wb)$. In 2012, using the integrated luminosity of 5.4 fb$^{-1}$, which is collected by the D0 Collaboration at the Tevatron $p\bar{p}$ Collider, $\Gamma_{t}=2.00_{-0.43}^{+0.47}$ GeV was extracted~\cite{D0:2012hgn}. In 2014, the CMS Collaboration provided a better determination of the total width, $\Gamma_{t}=1.36\pm 0.02(\rm stat.)_{-0.11}^{+0.14}(\rm syst.)$ GeV~\cite{CMS:2014mxl}, where ``stat." and ``syst." are short notations for statistical and systematic errors, respectively. This measurement is based on the assumption $\mathcal{B}(t\to Wq) = 1$, which includes the sum over all down-type quarks $q=(b,s,d)$. In 2017, an initial direct measurement was conducted by an ATLAS analysis, which involves the direct fitting of reconstructed lepton$+$jets events by using the integrated luminosity of 20.2 fb$^{-1}$ at a center-of-mass energy of $\sqrt{s}=8$ TeV. This resulted in $\Gamma_{t}=1.76\pm 0.33({\rm stat.})_{-0.68}^{+0.79}({\rm syst.})$ GeV~\cite{ATLAS:2017vgz}. In 2019, a measurement by the ATLAS Collaboration, using the integrated luminosity of 139 fb$^{-1}$ at the center-of-mass energy of $\sqrt{s}=13$ TeV, employed a template fit to the invariant mass of the lepton-$b$-quark in dilepton final states. This yielded $\Gamma_{t}=1.94_{-0.49}^{+0.52}$ GeV~\cite{ATLAS:2019onj}. The Particle Data Group (PDG) reported the world average as $\Gamma(t\to Wq)=1.42_{-0.15}^{+0.19}$ GeV and $\mathcal{B}(t\to Wb)=\Gamma(t\to Wb)/\Gamma(t\to Wq)=0.957\pm 0.034$~\cite{Workman:2022ynf}.

Theoretically, the next-to-leading order~(NLO) quantum chromodynamics~(QCD) corrections were first computed in Refs.\cite{Jezabek:1988iv, Czarnecki:1990kv, Li:1990qf, Jezabek:1993wk}, while the NLO EW corrections were provided in Refs.\cite{Denner:1990ns, Eilam:1991iz}. The next-to-next-to-leading order~(N$^{2}$LO) QCD corrections for the $t\to Wb$ decay had been done by using the asymptotic expansion~\cite{Czarnecki:1998qc, Chetyrkin:1999ju, Blokland:2004ye, Blokland:2005vq, Czarnecki:2001cz}, and the complete N$^{2}$LO analytical results were available in Ref.\cite{Chen:2022wit}. The N$^{2}$LO polarized decay rates were calculated in Refs.\cite{Czarnecki:2010gb, Czarnecki:2018vwh}. Recently, the next-to-next-to-next-to-leading order~(N$^{3}$LO) corrections in the large-$N_{C}$ limit have been presented in Ref.\cite{Chen:2023dsi}, and the first complete high-precision numerical results of N$^{3}$LO QCD corrections have also been given in Ref.\cite{Chen:2023osm}. Those two predictions agree well with each other, indicating that the leading-color contributions are dominant and the approximation of large-$N_{C}$ limit is highly reliable at least for this particular process.

Due to the large kinematic scale $Q \sim {\cal O}(m_{t})$ and the small strong coupling constant $\alpha_s(m_t)\sim 0.1$, the pQCD series for the $t\to Wb$ total decay width up to the N$^{3}$LO-level exhibits good convergence. It however still has a sizable renormalization scale dependence due to the divergent renormalon terms~\cite{Beneke:1994qe, Neubert:1994vb, Beneke:1998ui}. Practically, one usually selects the renormalization scale as $\mu_{R}=m_t$ so as to eliminate the divergent large logarithmic terms like $\ln \left(\mu_{R}^{2}/m_t^{2}\right)$, then vary it within a certain range such as $\mu_{R}\in [m_t/\xi, \xi m_t]$ with $\xi$ being usually chosen as $2$, $3$, $4$, and etc., to account for its uncertainty. This treatment is referred to as the conventional scale-setting approach. It is evident that this approach is arbitrary, and the perturbative nature of the series heavily relies on the choice of $\xi$, thereby diminishing the reliability of the final theoretical prediction. As is well-known, the physical observable, which corresponds to an infinite-order perturbative series, should be scale invariant~\cite{Petermann:1953wpa, Peterman:1978tb, Callan:1970yg, Symanzik:1970rt}; and it is important to known whether such scale invariance can also be achieved for a fixed-order series. Simply requiring the fixed-order series to be scale invariant does not achieve this goal, as it is not true and explicitly breaks standard renormalization group invariance (RGI)~\cite{Wu:2014iba}. Since this simple requirement implicitly assumes that all uncalculated higher-order terms contribute zero. Therefore, a proper method needs to be introduced to improve the perturative series before applying the scale-setting procedures.

Many scale-setting approaches have been suggested to improve the fixed-order series so as to achieve a scale invariant prediction. Especially, by using the $n_f$-terms as a guide, the Brodsky--Lepage--Mackenzie~(BLM) approach~\cite{Brodsky:1982gc} automatically resums the corresponding gluons as well as the quark vacuum-polarization contributions, which then leads to a scheme-and-scale invariant prediction~\cite{Brodsky:1994eh}. The NLO commensurate scale relations which ensure the scheme-and-scale invariance at the NLO level have also been given there. Those relations indicate that if the expansion coefficients match well with the corresponding $\alpha_s$, exactly scheme-independent predictions can be achieved. Since the running behavior of $\alpha_{s}$ is governed by the renormalization group equation~(RGE) or the $\beta$-function~\cite{Gross:1973ju, Politzer:1974fr}, to deal with the $\{\beta_i\}$-terms involved in the RGE is then more fundamental than to deal with the $n_f$-terms. Lately, the BLM is developed to Principle of Maximum Conformality~(PMC)~\cite{Brodsky:2011ta, Brodsky:2011ig, Mojaza:2012mf, Brodsky:2012rj, Brodsky:2013vpa}, which perfects the idea behind BLM and offers a reliable extension of the BLM to all orders. In the Abelian limit~\cite{Brodsky:1997jk}, BLM and PMC reduce to the well-known Gell-Mann-Low approach~\cite{Gell-Mann:1954yli} for QED. Subsequently, the PMC single-scale-setting approach~(PMCs), as an effective alternative to the original PMC multi-scale-setting approach, has also been proposed in Refs.\cite{Shen:2017pdu, Yan:2022foz} from two distinct but equivalent perspectives. It has been demonstrated that the PMC prediction is independent to any choice of renormalization scheme and scale~\cite{Wu:2018cmb}, being consistent with the self-consistency requirements of the renormalization group~\cite{Brodsky:2012ms, Wu:2019mky}. The PMCs approach also greatly suppresses the residual scale dependence~\cite{Zheng:2013uja} of the PMC predictions due to unknown even higher-order terms of the pQCD series. In this paper, we adopt the PMCs approach to deal with the $t\to Wb$ total decay width up to N$^{3}$LO QCD corrections.

For any perturbative series, there are some uncertainties caused by the unknown higher-order terms~(UHO-terms). Since the exact pQCD result is unknown, it would be helpful to quantify the UHO-terms' contribution in terms of a probability distribution. Following the idea of Bayesian analysis (BA)~\cite{Cacciari:2011ze, Bagnaschi:2014wea, Bonvini:2020xeo, Duhr:2021mfd}, the conditional probability of the unknown perturbative coefficient is first given by a subjective prior distribution, which is then updated iteratively according to the Bayes' theorem as more and more information has been included. It has been found that the generally more convergent and scheme-and-scale invariant PMC series provides a more reliable basis than conventional series for estimating the contributions from the UHO-terms~\cite{Du:2018dma, Shen:2022nyr, Shen:2023qgz, Luo:2023cpa}. In this paper, we adopt the BA approach to estimate the contributions from the unknown N$^4$LO-terms of the PMC series.

The remaining parts of the paper are organized as follows. Sec.\ref{sec2} gives the formulas for the top-quark decay process, $t\to Wb$, up to N$^{3}$LO QCD corrections, and then shows how to apply the PMC scale-setting procedures to the present process. Sec.\ref{sec3} presents numerical results and discussions for the top-quark decay.  Sec.\ref{sec4} is reserved for a summary.

\section{calculation technology}\label{sec2}

By neglecting the corrections caused by the finite $b$-quark mass and the off-shell $W$-boson effect, the total decay width of the top-quark decay $t\to Wb$ up to N$^{3}$LO QCD corrections can be expressed as
\begin{align} \label{Series0}
\Gamma_{t\to Wb}^{\rm QCD} = \frac{G_{F} m_{t}^{3} |V_{\rm tb}|^{2}} {8\sqrt{2}\pi}\sum_{i=0}^{3}f_{i}(\omega;\mu_{R}) \alpha_{s}^{i}(\mu_{R}),
\end{align}
where $\mu_R$ is the renormalization scale, $G_{F}$ is the Fermi constant, $m_{t}$ is the top-quark pole mass, $V_{\rm tb}$ is the Cabibbo-Kobayashi-Maskawa~(CKM) matrix element, and $\omega = m_{W}^{2}/m_{t}^{2}$. The LO decay width that is caused by weak interaction provides dominant contribution to the total decay width. Because the $W$-boson will decay promptly into leptons or quarks, we rewrite the top-quark decay width as
\begin{align} \label{gammaQCD}
\Gamma_{t\to W^{*}b}^{\rm QCD}  = \Gamma_{\rm LO}\left[1 + R_{t}(\mu_{R})\right],
\end{align}
where $W^*$ indicates the $W$-boson may be off-shell and the QCD corrections
\begin{align}\label{Rt}
R_{t}(\mu_{R}) = \sum_{i=1}^{3}r_{i}(\mu_{R})\alpha_{s}^{i}(\mu_{R})+{\cal O}(\alpha_{s}^{4}),
\end{align}
where for the coefficients $r_i$ ($i\in[1,3]$), we have~\cite{Jezabek:1988iv, Chen:2022wit}
\begin{align}\label{coeff}
	r_{i}(\mu_{R}) = \frac{1}{\Gamma_{\rm LO}}\frac{G_{F} m_{t}^{3} |V_{\rm tb}|^{2}} {8\sqrt{2}\pi}
    \frac{\omega\gamma}{\pi}\int_{0}^{\bar{\varepsilon}} \frac{f_{i}(\varepsilon,x;\mu_{R})}{(x-\omega)^{2}+\omega^{2}\gamma^{2}} \, {\rm d}x.
\end{align}
Here $\varepsilon = m_{b}^{2}/m_{t}^{2}$, $\bar{\varepsilon}=(1-m_{b}/m_{t})^{2}$ and $\gamma = \Gamma_{W}/m_{W}$ with $\Gamma_{W}$ being the $W$-boson total decay width. The LO decay width is scale invariant and equals to
\begin{align}\label{gammaLO}
	\Gamma_{\rm LO} = \frac{G_{F} m_{t}^{3} |V_{\rm tb}|^{2}} {8\sqrt{2}\pi} \frac{\omega\gamma}{\pi}
    \int_{0}^{\bar{\varepsilon}} \frac{f_{0}(\varepsilon,x)}{(x-\omega)^{2}+\omega^{2}\gamma^{2}} \, {\rm d}x.
\end{align}
To get the analytic expression, the integration involving the polylogarithm functions can be accomplished with the assistance of the \texttt{PolyLogTools}~\cite{Duhr:2019tlz} and \texttt{GiNaC}~\cite{Bauer:2000cp} packages. In Eqs.(\ref{coeff}, \ref{gammaLO}), the finite $b$-quark mass contributions have been included, e.g. the coefficients $f_{i}(\omega;\mu_{R})$ in Eq.\eqref{gammaQCD} have been replaced by $f_{i}(\varepsilon,\omega;\mu_{R})$ and the upper limit of the integral in Eqs.(\ref{coeff}, \ref{gammaLO}) becomes $\bar{\varepsilon}$ instead of $1$. Currently, the finite $b$-quark mass effect is known only up to NLO accuracy~\cite{Jezabek:1988iv}. For convenience, we list the analytic coefficients $f_{0}$ and $f_{1}$ at the scale $\mu_R=m_t$, including and excluding $b$-quark mass effects, in the Appendix. The analytic form of the coefficients $f_{2}$ and $f_{3}$ can be found in Refs.\cite{Chen:2022wit, Chen:2023dsi}, where $f_{3}$ contains only the most dominant leading color contribution that give the most dominant contributions.

Following the standard PMC single-scale setting procedures~\cite{Shen:2017pdu, Yan:2022foz}, by using the QCD degeneracy relations among different orders~\cite{Bi:2015wea}, one can distribute the original $n_f$-series into conformal and non-conformal terms, and then use the RGE-involved non-conformal $\{\beta_i\}$-terms to determine an overall effective running coupling $\alpha_{s}(Q_{*})$ for $t\to Wb$ decay, where $Q_{*}$ is referred to as the PMC scale, which can be viewed as the effective momentum flow for the process. More explicitly, the coefficients $r_{i}$ in Eq.\eqref{gammaQCD} can be parameterized as follows:
\begin{align}
	r_{1} & = r_{1,0},\\
	r_{2} & = r_{2,0} + \beta_{0} r_{2,1},\\
	r_{3} & = r_{3,0} + \beta_{1} r_{2,1} + 2\beta_{0} r_{3,1} + \beta_{0}^{2} r_{3,2}.
\end{align}
Here, $r_{i,0}$ are scale-invariant conformal coefficients and $r_{i,j(\neq 0)}$ are non-conformal coefficients. At present, the $\{\beta_i\}$-functions have been computed up to the five-loop level in the modified minimal-subtraction scheme~($\overline{\rm MS}$-scheme)~\cite{Baikov:2016tgj, Herzog:2017ohr}, e.g. $\beta_{0}=\left(11-2n_{f}/3\right)/(4\pi)$ and $\beta_{1}=\left(102-38n_f/3\right)/(4\pi)^2$ for $n_f$ active flavors.

After applying the PMC, the divergent renormalon terms, which grow as various powers of $\beta_0$ under the approximation $\beta_i\simeq \beta_0^{i+1}$, have been removed, we then obtain a more convergent pQCD series as
\begin{align}\label{PMCseries}
	R_{t}\big|_{\rm PMC}=\sum_{i=1}^{3}r_{i,0}\alpha_{s}^{i}(Q_{*})+{\cal O}(\alpha_{s}^{4}),
\end{align}
where $Q_{*}$ is fixed by requiring all non-conformal terms to vanish. Using the present known pQCD series up to N$^{3}$LO level, the PMC scale $Q_{*}$ can be fixed up to next-to-leading-log (NLL) accuracy, i.e.,
\begin{align}\label{Qs}
	\ln\frac{Q_{*}^{2}}{Q^{2}}=S_{0}+S_{1}\alpha_{s}(Q_{*})+{\cal O}\left(\alpha_{s}^{2}\right),
\end{align}
where the coefficients $S_{0,1}$ are
\begin{align}
	S_{0} & = -\frac{\hat{r}_{2,1}}{\hat{r}_{1,0}},\label{S1}\\
	S_{1} & = \frac{2(\hat{r}_{2,0}\hat{r}_{2,1}-\hat{r}_{1,0}\hat{r}_{3,1})}{\hat{r}_{1,0}^{2}} +\frac{\hat{r}_{2,1}^{2}-\hat{r}_{1,0}\hat{r}_{3,2}}{\hat{r}_{1,0}^{2}}\beta_{0},\label{S2}
\end{align}
where $\hat{r}_{i,j}\equiv r_{i,j}|_{\mu_{R}=Q}$. Since all the non-conformal $\{\beta_{i}\}$-terms of the series (\ref{Rt}) that are associated with the RGE have been absorbed into $\alpha_{s}(Q_{*})$, the resulting pQCD series~\eqref{PMCseries} becomes conformal series. The magnitude of $\alpha_{s}$ then matches well with the expansion coefficients of series, yielding naturally scheme-independent theoretical predictions at any fixed order~\cite{Wu:2013ei, Wu:2014iba, Wu:2018cmb, Wu:2019mky, DiGiustino:2023jiq}.

\section{Numerical results and discussions}
\label{sec3}

For numerical calculations, the following values are taken as the input parameters~\cite{Workman:2022ynf}:
\begin{align}
	m_{b}&=4.78\ {\rm GeV},&m_{t}&=172.5 \pm 0.7\ {\rm GeV},\notag\\
	m_{W}&=80.377\ {\rm GeV},&\Gamma_{W}&=2.085\pm 0.042\ {\rm GeV},\notag\\
	m_{Z}&=91.1876\ {\rm GeV},&G_{F}&=1.1663788\times10^{-5}\ {\rm GeV}^{2}, \notag
\end{align}
and $\alpha_s(m_{Z})=0.1179\pm 0.0009$. The NLO EW correction $\Delta_{\rm EW}$ can be calculated using the formulas given in Refs.\cite{Denner:1990ns, Eilam:1991iz, Denner:1990cpz} together with the recently issued Mathematica program \texttt{TopWidth}~\cite{Chen:2022wit, Chen:2023dsi}. By using the above inputs, we obtain $\Delta_{\rm EW}=0.0249\pm 0.0004$ GeV, where the error are for $\Delta m_{t}=\pm 0.7$ GeV.

\subsection{Basic properties}

Using Eqs.(\ref{Qs},\ref{S1},\ref{S2}), we can obtain the wanted LL-accuracy and NLL-accuracy PMC scales that correspond to N$^{2}$LO and N$^{3}$LO pQCD series, respectively, e.g.,
\begin{align}
	Q_{*}^{(\rm LL, NLL)}=\{15.2143, 10.3048\}\ {\rm GeV}.
\end{align}
Using the NLL-accuracy $Q_{*}$, we then obtain the N$^3$LO-level pQCD approximant $R_{t}$ under the PMC scale-setting approach
\begin{equation}
R_{t}\big|_{\rm PMC}  = -0.1146. \label{numRtPMC}
\end{equation}
We observe that the improved pQCD series after applying the PMC becomes independent to any choice of renormalization scale. On the other hand, we observe that the scale dependence of the original pQCD series does become smaller when more loop terms have been included, being consistent with conventional wisdom. As for the present N$^3$LO-level series, the net error under conventional scale-setting approach is small which is about $2.5\%$ for $\mu_{R}\in [m_{t}/2, 2m_{t}]$, which will be extended to $\simeq 4.8\%$ for a broader choice of scale range $\mu_{R}\in [m_{t}/4, 4m_{t}]$. More explicitly, we give the N$^3$LO-level conventional prediction in the following,
\begin{equation}
R_{t}\big|_{\rm Conv.} = -0.1122_{-0.0019-0.0013}^{+0.0009+0.0014}, \label{numRtConv}
\end{equation}
whose central value corresponds to $\mu_{R}=m_{t}$, the first errors are for $\mu_{R}\in [m_{t}/2, 2m_{t}]$ and the second ones are additional errors by using a broader range $\mu_{R}\in [m_{t}/4, 4m_{t}]$. The net N$^3$LO $R_{t}$ for conventional and PMC series are consistent with each other under proper choice of scale range for conventional scale-setting approach. Thus we need to be careful of discussing the scale uncertainties under conventional scale-setting approach. For convenience, if not specially stated, we will adopt the usual choice of $\mu_{R}\in [m_{t}/2, 2m_{t}]$ to do our discussions. It is found that the perturbative behavior of conventional series still depends heavily on the choice of scale. To show this point more clearly, we define a $\kappa$-factor for the series (\ref{Rt}) or (\ref{PMCseries}), i.e.,
\begin{align}
\kappa_{\rm Conv.}^{\rm N^{\it i}LO} &= \frac{r_{i}(\mu_{R})\alpha_{s}^{i}(\mu_{R})}{r_{1}\alpha_{s}(\mu_{R})}, &\kappa_{\rm PMC}^{(i)} &= \frac{r_{i,0}\alpha_{s}^{i}(Q_{*})}{r_{1,0}\alpha_{s}(Q_{*})}.
\end{align}
Numerically, we have
\begin{align}
	\kappa_{\rm Conv.} &= \{1,0.1702,0.0476\},\quad \mu_{R} = m_{t}/2  \label{kappa1}  \\
	\kappa_{\rm Conv.} &= \{1,0.2450,0.0792\},\quad \mu_{R} = m_{t}    \label{kappa2}  \\
	\kappa_{\rm Conv.} &= \{1,0.3096,0.1155\},\quad \mu_{R} = 2m_{t}   \label{kappa3}  \\
	\kappa_{\rm PMC}   &= \{1,0.1210,0.0548\}. \label{kappa4}
\end{align}
The scale-independent convergent behavior of the PMC series (\ref{PMCseries}) could be regarded as the intrinsic perturbative nature of $R_t$. According to Eqs.(\ref{numRtPMC}, \ref{numRtConv}), after including the known NLO EW correction $\Delta_{\rm EW}$, the total top-quark decay width of $t\to Wb$ are
\begin{align}
	\Gamma_{t}^{\rm tot}\big|_{\rm Conv.} & = 1.3156_{-0.0027}^{+0.0014}\ {\rm GeV},\\
	\Gamma_{t}^{\rm tot}\big|_{\rm PMC}   & = 1.3120\ {\rm GeV},
\end{align}
where the errors for the series under conventional scale-setting approach is for $\mu_{R}\in [m_{t}/2, 2m_{t}]$.

\begin{figure} [htb]
\centering
\includegraphics[width=0.48\textwidth]{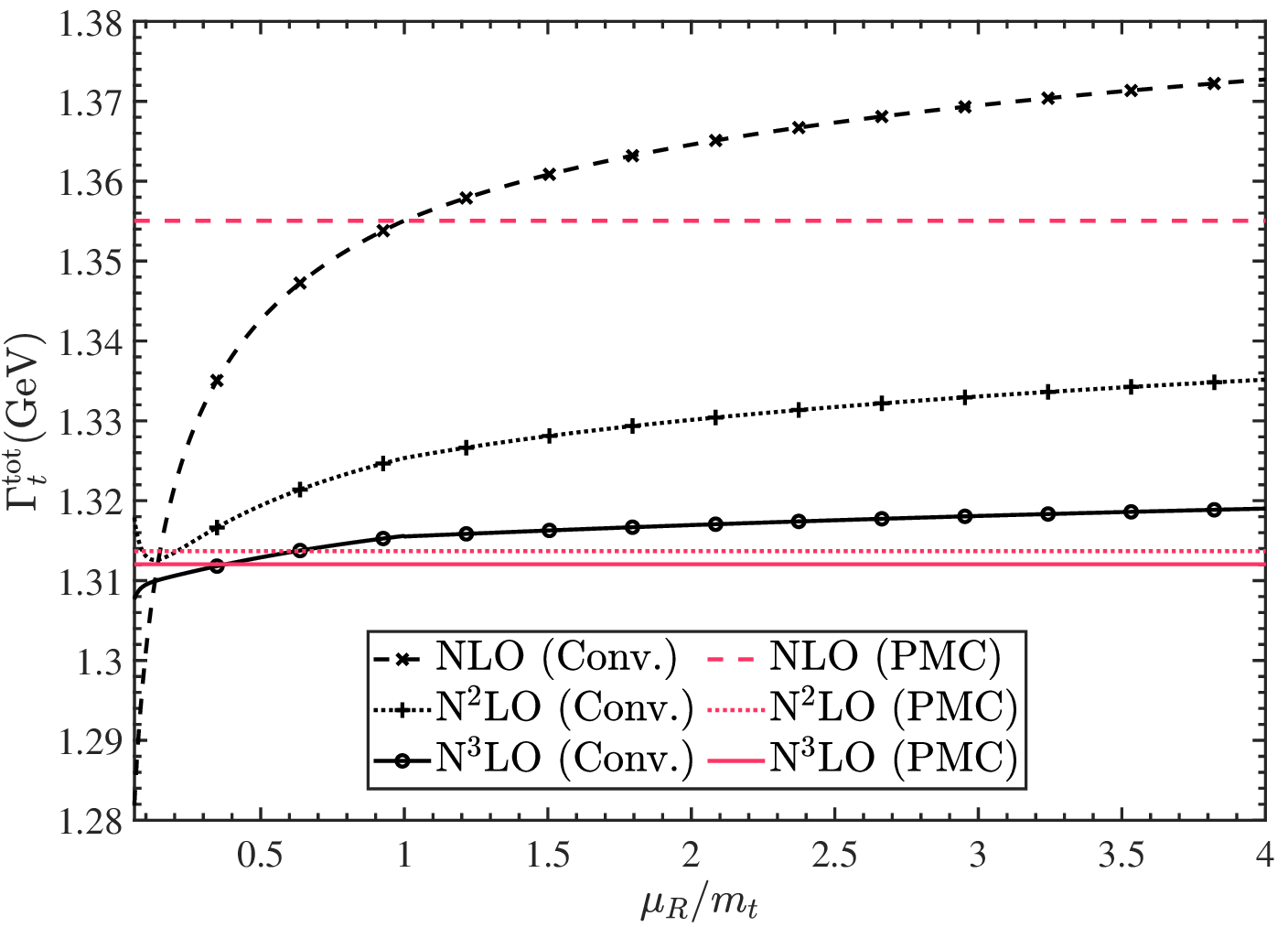}
\caption{The total top-quark decay width of $t\to Wb$ ($\Gamma_{t}^{\rm tot}$) up to N$^3$LO QCD corrections versus the renormalization scale $\mu_{R}$ using conventional (Conv.) (black lines) and PMC (red lines) scale-setting approaches, respectively. }
\label{topdecayConv}
\end{figure}

We present the top-quark total decay width $\Gamma_{t}^{\rm tot}$ up to N$^3$LO QCD corrections versus the renormalization scale $\mu_{R}$ (divided by $m_t$) before and after applying the PMC scale-setting approach in FIG.\ref{topdecayConv}. After applying the PMC, as shown in FIG.\ref{topdecayConv}, the scale dependence of the top-quark decay width is eliminated at any fixed-order. The difference between N$^2$LO-level and the N$^3$LO-level PMC predictions are much smaller than that of the conventional ones, indicating that the convergence of the pQCD series is significantly improved.

\subsection{Predictions of the uncalculated N$^4$LO contributions using the Bayesian analysis approach}

It has been noted that the improved series by using the PMC scale-setting approach not only provides more precise predictions for the fixed-order pQCD series but also establishes a robust basis for estimating the potential contributions from the UHO-terms, thus significantly improving the predictive power of perturbation theory.

In the following, we adopt the BA to estimate the effects from the uncalculated N$^4$LO-terms from the known initial N$^{3}$LO-level series~\eqref{Rt} and the PMC series~\eqref{PMCseries}, respectively. The BA quantifies the contributions of the UHO-terms in terms of a probability distribution. It becomes most effective when the series has a good convergent behavior. A detailed introduction to the BA can be found in Refs.~\cite{Cacciari:2011ze, Bagnaschi:2014wea, Bonvini:2020xeo, Duhr:2021mfd} and its combination with the PMC can be found in Refs.\cite{Du:2018dma, Shen:2022nyr, Yan:2022foz, Shen:2023qgz, Luo:2023cpa}. Following the idea of the BA, for a fixed degree-of-belief (DoB) or equivalently the Bayesian probability, the estimated UHO-coefficient $r_{p+1}$, given known coefficients $\{r_{1},r_{2},\cdots,r_{p}\}$, will fall within the following specific credible interval~(CI) $r_{p+1}\in\left[-r_{p+1}^{(\rm DoB)},r_{p+1}^{(\rm DoB)}\right]$, where
\begin{align}
	r_{p+1}^{(\rm DoB)}=\left\{
	\begin{aligned}
		&\bar{r}_{(p)}\frac{p+1}{p}{\rm DoB},&{\rm DoB}&\le \frac{p}{p+1}\\
		&\bar{r}_{(p)}\left[(p+1)(1-{\rm DoB})\right]^{-1/p},&{\rm DoB}&\ge \frac{p}{p+1}
	\end{aligned}\right.
\end{align}
with $\bar{r}_{(p)}={\rm max}\{|r_{1}|,|r_{2}|,\cdots,|r_{p}|\}$. For definiteness and without loss of generality, we take ${\rm DoB}\equiv 95.5\%$ to estimate the contributions from the UHO-terms.

\begin{figure} [htb]
\centering
\includegraphics[width=0.47\textwidth]{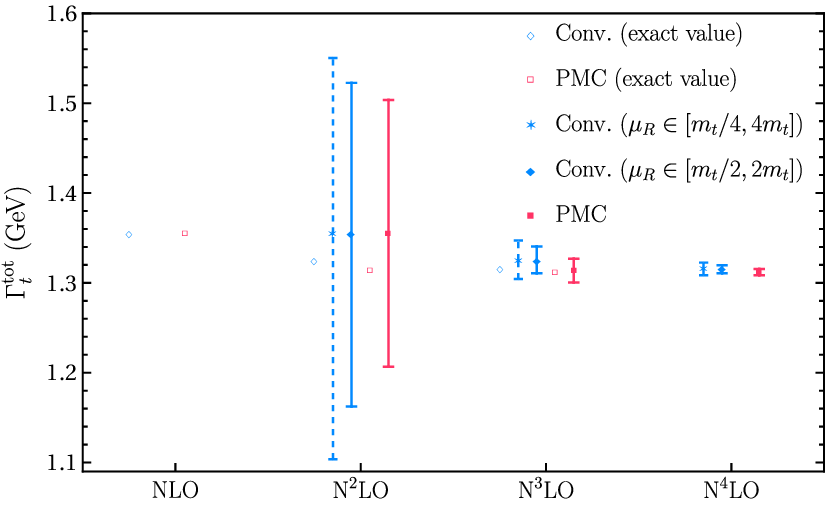}
\caption{Comparison of the calculated central values of the total top-quark decay width of $t\to Wb$ ($\Gamma_{t}^{\rm tot}$) for the known series (labeled as ``exact value'') with the predicted credible intervals of $\Gamma_{t}^{\rm tot}$ up to N$^4$LO-level QCD corrections. The blue hollow diamonds and red hollow quadrates represent the calculated central values of the fixed-order pQCD predictions using conventional (Conv.) and PMC scale-setting approaches, respectively. The blue stars and blue solid diamonds and red solid quadrates with error bars represent the predicted credible intervals using the Bayesian approach based on the known Conv. series and the PMC series, respectively. ${\rm DoB}=95.5\%$. }
\label{UHOerr}
\end{figure}

Comparison of the calculated central values using known series (simply labeled as ``exact value'') of the total top-quark decay width $\Gamma_{t}^{\rm tot}$ with the predicted credible intervals of $\Gamma_{t}^{\rm tot}$ up to N$^4$LO-level QCD corrections are given in FIG.\ref{UHOerr}. Due to the scale-dependence of the coefficients $r_{i>1}$ in the conventional pQCD series, the BA can only be employed after specifying the renormalization scale, which introduces additional uncertainties to the total decay width. And in FIG.\ref{UHOerr}, we give two predictions for $\mu_{R}\in [m_{t}/2, 2m_{t}]$ and $\mu_{R}\in [m_{t}/4, 4m_{t}]$, respectively. In contrast, the conformal coefficients $r_{i,0}$ in the PMC series are scale-independent, providing a more reliable foundation for constraining estimations from UHO contributions. FIG.\ref{UHOerr} shows that the probability distribution become more accurate, simultaneously the resultant credible intervals become smaller for the same DoB, when more loop terms have been included. Practically, we can treat the magnitude of unknown N$^4$LO-terms as one of the errors of the given N$^3$LO-level $\Gamma_{t}^{\rm tot}$. If taking the scale range $\mu_{R}\in [m_{t}/2, 2m_{t}]$ to estimate the UHO-contributions, such errors by using the BA are
\begin{align}
	\Delta\Gamma_{t}^{\rm tot}\big|_{\rm Conv.}^{\rm UHO} &=\left(_{-0.0041}^{+0.0037}\right)\ {\rm GeV},\\
	\Delta\Gamma_{t}^{\rm tot}\big|_{\rm PMC}^{\rm UHO}&= \pm0.0035\ {\rm GeV}.
\end{align}
The predicted error range for the perturbative series under conventional scale-setting approach will be extended to $\Delta\Gamma_{t}^{\rm tot}\big|_{\rm Conv.}^{\rm UHO} =\left(_{-0.0054}^{+0.0060}\right)\ {\rm GeV}$ for $\mu_{R}\in [m_{t}/4, 4m_{t}]$. The above two equations show that after applying the PMC, the errors $\Delta\Gamma_{t}^{\rm tot}$ from the UHO-terms are only slightly smaller than those of the pQCD series under conventional scale-setting approach. Both of which are small, e.g. $\kappa^{\rm N^{4}LO}\sim 0.03$ for all cases, due to the fact that the N$^3$LO-level QCD corrections already show good convergent behavior as indicated by Eqs.(\ref{kappa1}-\ref{kappa4}).

\subsection{Uncertainties for the total decay width $\Gamma_{t}^{\rm tot}$}

In addition to the above mentioned scale uncertainties and the uncertainties caused by the UHO-terms, there are also other error sources for the determination of $t\to Wb$ decay width. When discussing the uncertainty for one of the error sources, the other error sources are set to be their central values.

\begin{table}[htbp]
\centering
\begin{tabular}{cccc}
\hline
		& ~~$\Delta\Gamma_{t}^{\rm tot}\big|^{\Delta\alpha_{s}}$~~ & ~~$\Delta\Gamma_{t}^{\rm tot}\big|^{\Delta\Gamma_{W}}$~~ & ~~$\Delta\Gamma_{t}^{\rm tot}\big|^{\Delta m_{t}}$~~ \\
\hline
	~~Conv.~~ & $\pm 0.0015$ & $\pm 0.0004$ & $\left(_{-0.0189}^{+0.0190}\right)$ \\
\hline
	~~PMC~~ & $\pm 0.0015$ & $\pm 0.0004$ & $\left(_{-0.0188}^{+0.0190}\right)$ \\
\hline
\end{tabular}
\caption{Additional uncertainties (in unit: GeV) arising from $\Delta \alpha_{s}(m_{Z})=\pm 0.0009$, $\Delta \Gamma_{W}= \pm 0.042$ GeV, and $\Delta m_{t}= \pm 0.7$ GeV under conventional and PMC scale-setting approaches, respectively. } \label{err}
\end{table}

We put the uncertainties arising from $\Delta \alpha_{s}(m_{Z})=\pm 0.0009$, $\Delta \Gamma_{W}= \pm 0.042$ GeV, and $\Delta m_{t}= \pm 0.7$ GeV under conventional and PMC scale-setting approaches in TABLE~\ref{err}. It shows that the errors are dominated by $\Delta m_{t}$, whose effect to the total decay width is about five to ten times larger than those of other error sources. Thus a more precise $m_t$ will greatly improve the precision of the theoretical predictions. The squared average of the uncertainties to the total top-quark decay width arising from $\mu_R\in[m_t/2, 2m_t]$, $\Delta \alpha_{s}(m_{Z})$, $\Delta \Gamma_{W}$, and the predicted N$^4$LO-terms are
\begin{align}
	\Gamma_{t}^{\rm tot}\big|_{\rm Conv.} &= 1.3156_{-0.0051}^{+0.0042} \ {\rm GeV}, \\
	\Gamma_{t}^{\rm tot}\big|_{\rm PMC}   &= 1.3120 \pm 0.0038 \ {\rm GeV}.
\end{align}
And if further including the uncertainty caused by $\Delta m_t$, we finally get
\begin{align}
	\Gamma_{t}^{\rm tot}\big|_{\rm Conv.} &= 1.3156\pm 0.0195\ {\rm GeV}, \label{tote1} \\
	\Gamma_{t}^{\rm tot}\big|_{\rm PMC}   &= 1.3120_{-0.0192}^{+0.0194}\ {\rm GeV}.  \label{tote2}
\end{align}

\begin{figure} [htb]
\centering
\includegraphics[width=0.48\textwidth]{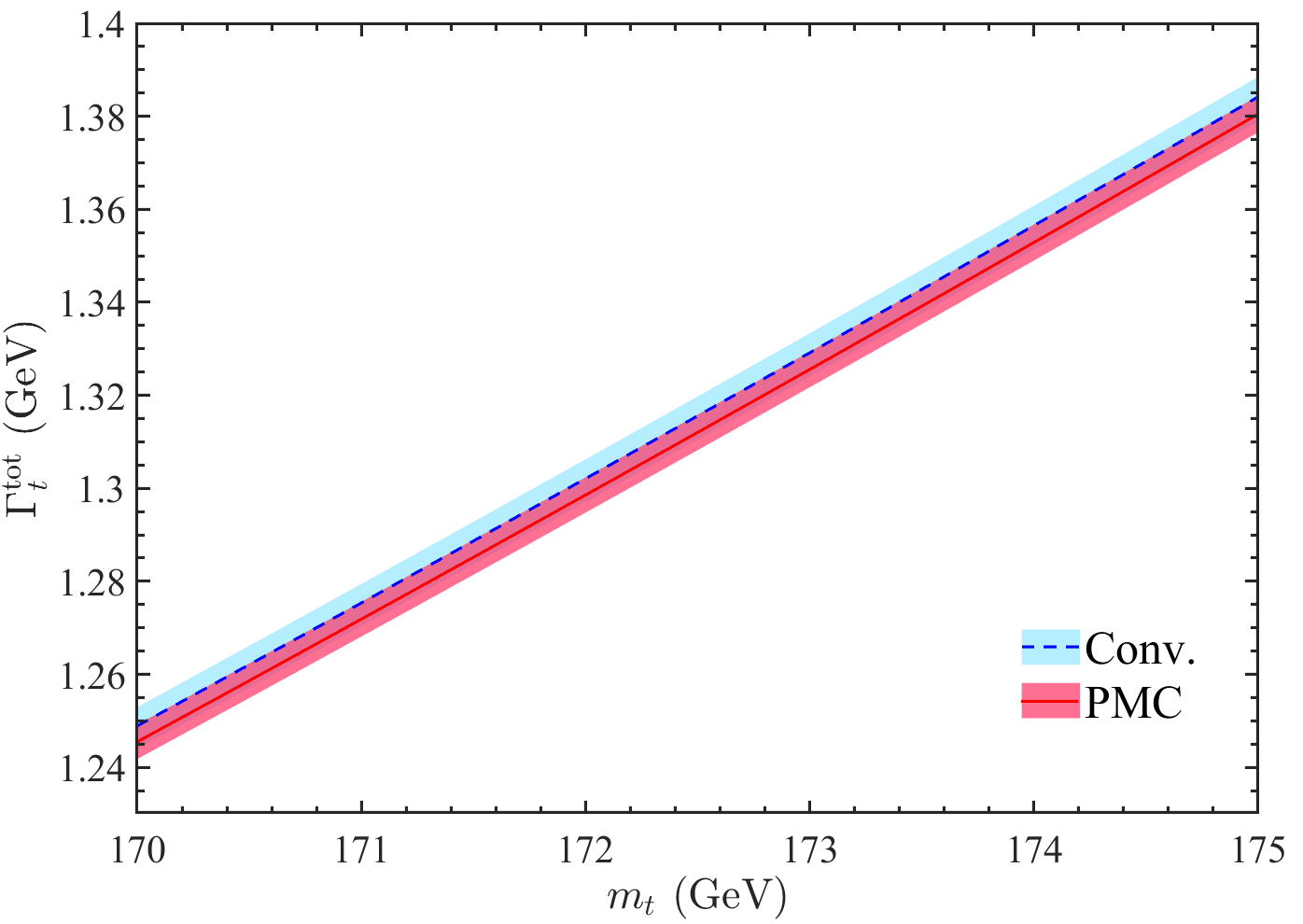}
\caption{Total decay width of  $t\to Wb$ ($\Gamma_{t}^{\rm tot}$) versus the top-quark pole mass $m_{t}$, where the light-blue and dark-red bands are results for conventional~(Conv.) and PMC scale-setting approaches, respectively. The dashed and solid lines are their central values. The shaded bands are for the uncertainties arising from $\mu_R\in[m_t/2, 2m_t]$, $\Delta \alpha_{s}(m_{Z})$, $\Delta \Gamma_{W}$, and the predicted N$^4$LO-terms. }
\label{gammamt}
\end{figure}

\begin{figure} [htb]
\centering
\includegraphics[width=0.48\textwidth]{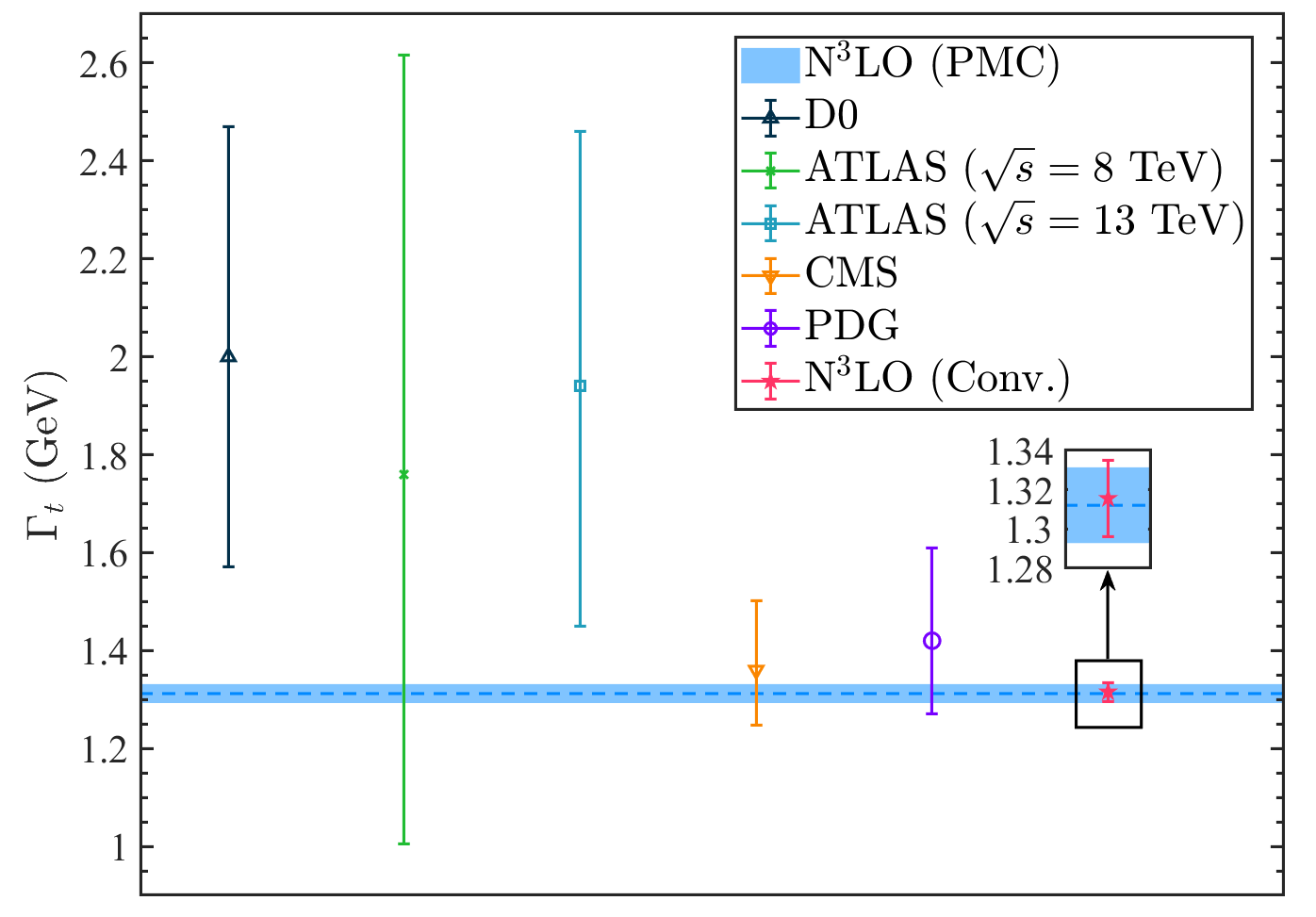}
\caption{Comparison between the theoretical predictions with their errors given in Eqs.(\ref{tote1}, \ref{tote2}) and the experimental measurements for the total decay width $\Gamma_{t}^{\rm tot}$. The errors of theoretical predictions are for the uncertainties arising from $\mu_R\in[m_t/2, 2m_t]$, $\Delta \alpha_{s}(m_{Z})$, $\Delta \Gamma_{W}$, $\Delta m_t$, and the predicted N$^4$LO-terms.}
\label{comparison}
\end{figure}

FIG.\ref{gammamt} depicts the relationship between the total decay width $\Gamma_{t}^{\rm tot}$ and the top-quark pole mass $m_{t}$, which indicates an approximately proportional relationship between the top-quark's decay widths and its pole mass. The total decay width is highly sensitive to the top-quark pole mass. FIG.\ref{comparison} shows the total decay width $\Gamma_{t}^{\rm tot}$ for the theoretical predictions given by Eqs.(\ref{tote1}, \ref{tote2}). As a comparison, various experimental measurements~\cite{D0:2012hgn, ATLAS:2017vgz, ATLAS:2019onj,CMS:2014mxl} and the world average value reported by the PDG~\cite{Workman:2022ynf} have been included in FIG.\ref{comparison}. Clearly, while the top quark's decay width is highly sensitive to its pole mass, the large uncertainties in experimental measurements preclude us from deriving a reliable reference value for the top-quark pole mass from these data.

In above analysis, we have implicitly taken the CKM matrix element $|V_{\rm tb}|=1$ to do our numerical calculation. As a final remark, we give an inverse determination of the CKM matrix element $|V_{\rm tb}|$ by using the PDG averaged values on the top-quark total decay width and the branching fraction $\mathcal{B}(t\to Wb)$. That's by using the PDG's averaged total decay width $\Gamma_{t,\rm PDG}=1.42_{-0.15}^{+0.19}$ GeV~\cite{Workman:2022ynf}, the branching fraction $\mathcal{B}(t\to Wb)=0.957\pm 0.034$, along with the theoretical predictions (\ref{tote1}, \ref{tote2}) under conventional and PMC scale-setting approaches, respectively, we inversely obtain~\footnote{The errors are estimated by using the usual error propagation formulas. That is, the error of a quantity $Z=X/Y$ is calculated by $\Delta Z = \left(X_{0}/Y_{0}\right)\sqrt{\left(\Delta X/X_{0}\right)^{2}+\left(\Delta Y/Y_{0}\right)^{2}}$, where $X = X_{0} \pm \Delta X$, $Y = Y_{0} \pm \Delta Y$ and $Z = \left(X_{0}/Y_{0}\right) \pm \Delta Z$.}
\begin{align}
	|V_{\rm tb}|_{\rm Conv.} &= 1.033_{-0.116}^{+0.144}.   \\
	|V_{\rm tb}|_{\rm PMC} &= 1.036_{-0.116}^{+0.144}.
\end{align}
Both of them are consistent with the average value of the Tevatron and LHC results, e.g. $|V_{\rm tb}|=1.014\pm 0.029$~\cite{Workman:2022ynf}. By confining the prior within the SM region $[0,1]$, we then establish a lower limit of $|V_{\rm tb}|>0.917$ and $|V_{\rm tb}|>0.919$ for conventional and PMC scale-setting approaches, respectively.  \\

\section{Summary }  \label{sec4}

In the paper, we have presented an improved analysis of the total decay width of the top-quark decay $t\to Wb$ up to N$^{3}$LO QCD corrections by applying the PMC scale-setting approach. In contrast to previous literature~\cite{Meng:2022htg}, our present treatment is achieved after taking both the off-shell $W$-boson contributions and the finite $b$-quark mass effects into account. Because the N$^3$LO-level QCD corrections to the total decay width $\Gamma(t\to Wb)$ already show good convergent behavior as indicated by Eqs.(\ref{kappa1}-\ref{kappa4}), the predictions under conventional scale-setting approach are close to the PMC predictions. Especially because the errors are dominated by $\Delta m_t$, which dilute the great improvements on perturbative nature of the series by applying the PMC. However, the improved pQCD series after applying the PMC is independent to any choice of renormalization scale, which not only leads to a more precise prediction, but also provides a better basis for estimating the contributions from UHO-terms. The errors of the PMC series caused by $\Delta\alpha_{s}(m_{Z})$ and the predicted N$^4$LO-terms are comparable to each other, and there are also sizable renormalization scale error for conventional scale-setting approach. FIG.\ref{topdecayConv} indicates that the difference between N$^2$LO-level and the N$^3$LO-level PMC predictions are much smaller than the conventional one, indicating that the convergence of the pQCD series is significantly improved and the PMC prediction shows a quicker trends of approaching to its physical/measured value. Thus our present results emphasize the importance of using proper scale-setting approaches to achieve precise fixed-order pQCD predictions.

\hspace{1cm}

\noindent{\bf Acknowledgements}: This work was supported in part by the Chongqing Graduate Research and Innovation Foundation under Grant No.CYB23011 and No.ydstd1912, and by the Natural Science Foundation of China under Grant No.12175025 and No.12347101.

\hspace{1cm}

\section*{Appendix}

The first two scale-independent coefficients $f_{0}(\varepsilon,\omega)$ and $f_{1}(\varepsilon,\omega)$ at the scale $\mu_R=m_t$ for $t\to Wb$ which contain finite $b$-quark mass effects are~\cite{Jezabek:1988iv}
\begin{widetext}
	\begin{align}
		f_{0}(\varepsilon,\omega)  = & \lambda^{1/2}(1,\varepsilon,\omega) 	\left[(1-\varepsilon)^{2}+\omega(1+\varepsilon)-2\omega^{2}\right],\\
		f_{1}(\varepsilon,\omega)  = & -\frac{C_{F}}{2\pi}\Bigg\{
		\left[(1-\varepsilon)^{2}+\omega(1+\varepsilon)-2\omega^{2}\right] (1+\varepsilon-\omega)
			\Bigg[
				\pi^{2} + 2{\rm Li}_{2}(u_{\rm W}) - 2{\rm Li}_{2}(1-u_{\rm W}) - 4{\rm Li}_{2}(u_{q}) - 4{\rm	Li}_{2}(u_{q}u_{\rm W}) \notag\\
				&\, + \ln\frac{1-u_{q}}{\omega} \ln(1-u_{q})-\ln^{2}(1-u_{q}u_{\rm W}) + \frac{1}{4}\ln^{2}\frac{\omega}{u_{\rm W}} - \ln u_{\rm W} \ln\frac{(1-u_{q}u_{\rm W})^{2}}{1-u_{q}} - 2\ln u_{q} \ln\left[(1-u_{q})(1-u_{q}u_{\rm W})\right]
			\Bigg]\notag\\
			&\,-2f_{0}(\varepsilon,\omega)\left(\ln\omega+\frac{3}{2}\ln\varepsilon-2\ln\lambda(1,\varepsilon,\omega)\right) + 2 (1-\varepsilon)\left[(1-\varepsilon)^{2}+\omega(1+\varepsilon)-4\omega^{2}\right]\ln u_{\rm W}\notag\\
			&\, + \frac{1}{2}\bigg[ (3-\varepsilon + 11 \varepsilon^{2}-\varepsilon^{3})+\omega (6-12\varepsilon +2\varepsilon^{2})-\omega^{2}(21+5\varepsilon)+12\omega^{3}\bigg]\ln u_{q}\notag\\
			&\, + \frac{3}{2}\lambda^{1/2}(1,\varepsilon,\omega)(1-\varepsilon)(1+\varepsilon-\omega)\ln \varepsilon + \frac{1}{2} \lambda^{1/2}(1,\varepsilon,\omega) \left[-5 +22\varepsilon - 5\varepsilon^{2}-9\omega (1+\varepsilon) + 6\omega^{2}\right]
		\Bigg\},
	\end{align}
\end{widetext}
where $\omega = m_{W}^{2}/m_{t}^{2}$, $\varepsilon=m_{b}^{2}/m_{t}^{2}$, $C_F=4/3$ is SU$_c$(3) color factor, the K\"{a}llen function $\lambda(x,y,z)=x^{2}+y^{2}+z^{2}-2(xy+xz+yz)$, the polylogarithm function ${\rm Li}_{n}(z)=\sum_{k=1}^{\infty}z^{k}/k^{n}$. The functions $u_{q}$ and $u_{\rm W}$ are defined as
\begin{align}
	u_{q} & = \frac{1 + \varepsilon - \omega - \lambda^{1/2}(1,\varepsilon,\omega)}{1 + \varepsilon - \omega + \lambda^{1/2}(1,\varepsilon,\omega)}, \notag\\
	u_{\rm W} & = \frac{1 - \varepsilon + \omega - \lambda^{1/2}(1,\varepsilon,\omega)}{1 - \varepsilon + \omega + \lambda^{1/2}(1,\varepsilon,\omega)}.
\end{align}
In the case where the $b$-quark mass effects have been ignored, one has
\begin{align}
	f_{0}(\omega) \equiv&\, f_{0}(0,\omega) = (1-\omega)^{2}(1+2\omega),\\
	f_{1}(\omega)     = &\, -\frac{C_{F}}{2\pi}\Bigg\{f_{0}(\omega)\Bigg[\pi^{2}+2{\rm Li}_{2}(\omega)-2{\rm Li}_{2}(1-\omega)\Bigg]\notag\\
	&\,+ 2\omega(1+\omega)(1-2\omega)\ln\omega+(1-\omega)^{2}(5+4\omega)\ln(1-\omega)\notag\\
	&\,-\frac{1}{2}(1-\omega)\left(5+9\omega-6\omega^{2}\right)\Bigg\}.
\end{align}


\begin{thebibliography}{99}
	
\bibitem{D0:2012hgn}
V.~M.~Abazov \textit{et al.} [D0],
``An Improved determination of the width of the top quark,''
Phys. Rev. D \textbf{85}, 091104 (2012).

\bibitem{CMS:2014mxl}
V.~Khachatryan \textit{et al.} [CMS],
``Measurement of the ratio $\mathcal B(t \to Wb)/\mathcal B(t \to Wq)$ in pp collisions at $\sqrt{s}$ = 8 TeV,''
Phys. Lett. B \textbf{736}, 33 (2014).

\bibitem{ATLAS:2017vgz}
M.~Aaboud \textit{et al.} [ATLAS],
``Direct top-quark decay width measurement in the $t\bar{t}$ lepton+jets channel at $\sqrt{s}$=8 TeV with the ATLAS experiment,''
Eur. Phys. J. C \textbf{78}, 129 (2018).

\bibitem{ATLAS:2019onj}
[ATLAS],
``Measurement of the top-quark decay width in top-quark pair events in the dilepton channel at $\sqrt{s}=13$ TeV with the ATLAS detector,''
ATLAS-CONF-2019-038.

\bibitem{Workman:2022ynf}
R.~L.~Workman \textit{et al.} [Particle Data Group],
``Review of Particle Physics,''
PTEP \textbf{2022}, 083C01 (2022).

\bibitem{Jezabek:1988iv}
M.~Jezabek and J.~H.~Kuhn,
``QCD Corrections to Semileptonic Decays of Heavy Quarks,''
Nucl. Phys. B \textbf{314}, 1 (1989).

\bibitem{Czarnecki:1990kv}
A.~Czarnecki,
``QCD corrections to the decay t ---\ensuremath{>} W b in dimensional regularization,''
Phys. Lett. B \textbf{252}, 467 (1990).

\bibitem{Li:1990qf}
C.~S.~Li, R.~J.~Oakes and T.~C.~Yuan,
``QCD corrections to $t \to W^{+} b$,''
Phys. Rev. D \textbf{43}, 3759 (1991).

\bibitem{Jezabek:1993wk}
M.~Jezabek and J.~H.~Kuhn,
The Top width: Theoretical update,
Phys. Rev. D \textbf{48}, 1910 (1993);
[erratum: Phys. Rev. D \textbf{49}, 4970 (1994)].

\bibitem{Denner:1990ns}
A.~Denner and T.~Sack,
``The Top width,''
Nucl. Phys. B \textbf{358}, 46-58 (1991).

\bibitem{Eilam:1991iz}
G.~Eilam, R.~R.~Mendel, R.~Migneron and A.~Soni,
``Radiative corrections to top quark decay,''
Phys. Rev. Lett. \textbf{66}, 3105 (1991).

\bibitem{Czarnecki:1998qc}
A.~Czarnecki and K.~Melnikov,
``Two loop QCD corrections to top quark width,''
Nucl. Phys. B \textbf{544}, 520 (1999).

\bibitem{Chetyrkin:1999ju}
K.~G.~Chetyrkin, R.~Harlander, T.~Seidensticker and M.~Steinhauser,
``Second order QCD corrections to Gamma(t ---\ensuremath{>} W b),''
Phys. Rev. D \textbf{60}, 114015 (1999).

\bibitem{Blokland:2004ye}
I.~R.~Blokland, A.~Czarnecki, M.~Slusarczyk and F.~Tkachov,
``Heavy to light decays with a two loop accuracy,''
Phys. Rev. Lett. \textbf{93}, 062001 (2004).

\bibitem{Blokland:2005vq}
I.~R.~Blokland, A.~Czarnecki, M.~Slusarczyk and F.~Tkachov,
``Next-to-next-to-leading order calculations for heavy-to-light decays,''
Phys. Rev. D \textbf{71}, 054004 (2005);
[erratum: Phys. Rev. D \textbf{79}, 019901 (2009)].

\bibitem{Czarnecki:2001cz}
A.~Czarnecki and K.~Melnikov,
``Semileptonic b ---\ensuremath{>} u decays: Lepton invariant mass spectrum,''
Phys. Rev. Lett. \textbf{88}, 131801 (2002).

\bibitem{Chen:2022wit}
L.~B.~Chen, H.~T.~Li, J.~Wang and Y.~Wang,
``Analytic result for the top-quark width at next-to-next-to-leading order in QCD,''
Phys. Rev. D \textbf{108}, 054003 (2023).

\bibitem{Czarnecki:2010gb}
A.~Czarnecki, J.~G.~Korner and J.~H.~Piclum,
``Helicity fractions of W bosons from top quark decays at NNLO in QCD,''
Phys. Rev. D \textbf{81}, 111503 (2010).

\bibitem{Czarnecki:2018vwh}
A.~Czarnecki, S.~Groote, J.~G.~K\"orner and J.~H.~Piclum,
``NNLO QCD corrections to the polarized top quark decay $t(\uparrow) \to X_b+W^+$,''
Phys. Rev. D \textbf{97}, 094008 (2018).

\bibitem{Chen:2023dsi}
L.~B.~Chen, H.~T.~Li, Z.~Li, J.~Wang, Y.~Wang and Q.~f.~Wu,
``Analytic third-order QCD corrections to top-quark and semileptonic b---\ensuremath{>}u decays,''
Phys. Rev. D \textbf{109}, 7 (2024).

\bibitem{Chen:2023osm}
L.~Chen, X.~Chen, X.~Guan and Y.~Q.~Ma,
``Top-Quark Decay at Next-to-Next-to-Next-to-Leading Order in QCD,''
[arXiv:2309.01937 [hep-ph]].

\bibitem{Beneke:1994qe}
 M.~Beneke and V.~M.~Braun,
 Naive nonAbelianization and resummation of fermion bubble chains,
 Phys.\ Lett.\ B {\bf 348}, 513 (1995).

\bibitem{Neubert:1994vb}
 M.~Neubert,
 Scale setting in QCD and the momentum flow in Feynman diagrams,
 Phys.\ Rev.\ D {\bf 51}, 5924 (1995).

\bibitem{Beneke:1998ui}
 M.~Beneke,
 Renormalons,
 Phys.\ Rept.\ {\bf 317}, 1 (1999).

\bibitem{Petermann:1953wpa}
 A.~Petermann,
 Normalization of constants in the quanta theory,
 Helv.\ Phys.\ Acta {\bf 26}, 499 (1953).

\bibitem{Peterman:1978tb}
 A.~Peterman,
 Renormalization Group and the Deep Structure of the Proton,
 Phys.\ Rept.\ {\bf 53}, 157 (1979).

\bibitem{Callan:1970yg}
 C.~G.~Callan, Jr.,
 Broken scale invariance in scalar field theory,
 Phys.\ Rev.\ D {\bf 2}, 1541 (1970).

\bibitem{Symanzik:1970rt}
 K.~Symanzik,
 Small distance behavior in field theory and power counting,
 Commun.\ Math.\ Phys.\ {\bf 18}, 227 (1970).

\bibitem{Wu:2014iba}
X.~G.~Wu, Y.~Ma, S.~Q.~Wang, H.~B.~Fu, H.~H.~Ma, S.~J.~Brodsky and M.~Mojaza,
``Renormalization Group Invariance and Optimal QCD Renormalization Scale-Setting,''
Rept. Prog. Phys. \textbf{78}, 126201 (2015).

\bibitem{Brodsky:1982gc}
S.~J.~Brodsky, G.~P.~Lepage and P.~B.~Mackenzie,
``On the Elimination of Scale Ambiguities in Perturbative Quantum Chromodynamics,''
Phys. Rev. D \textbf{28}, 228 (1983).

\bibitem{Brodsky:1994eh}
 S.~J.~Brodsky and H.~J.~Lu,
 ``Commensurate scale relations in quantum chromodynamics'',
 Phys.\ Rev.\ D \textbf{51}, 3652 (1995).

\bibitem{Gross:1973ju}
D.~J.~Gross and F.~Wilczek,
``Asymptotically Free Gauge Theories - I,''
Phys. Rev. D \textbf{8}, 3633 (1973).

\bibitem{Politzer:1974fr}
H.~D.~Politzer,
``Asymptotic Freedom: An Approach to Strong Interactions,''
Phys. Rept. \textbf{14}, 129 (1974).

\bibitem{Brodsky:2011ta}
S.~J.~Brodsky and X.~G.~Wu,
``Scale Setting Using the Extended Renormalization Group and the Principle of Maximum Conformality: the QCD Coupling Constant at Four Loops,''
Phys. Rev. D \textbf{85}, 034038 (2012).

\bibitem{Brodsky:2012rj}
S.~J.~Brodsky and X.~G.~Wu,
``Eliminating the Renormalization Scale Ambiguity for Top-Pair Production Using the Principle of Maximum Conformality,''
Phys. Rev. Lett. \textbf{109}, 042002 (2012).

\bibitem{Brodsky:2011ig}
S.~J.~Brodsky and L.~Di Giustino,
``Setting the Renormalization Scale in QCD: The Principle of Maximum Conformality,''
Phys. Rev. D \textbf{86}, 085026 (2012).

\bibitem{Mojaza:2012mf}
M.~Mojaza, S.~J.~Brodsky and X.~G.~Wu,
``Systematic All-Orders Method to Eliminate Renormalization-Scale and Scheme Ambiguities in Perturbative QCD,''
Phys. Rev. Lett. \textbf{110}, 192001 (2013).

\bibitem{Brodsky:2013vpa}
S.~J.~Brodsky, M.~Mojaza and X.~G.~Wu,
``Systematic Scale-Setting to All Orders: The Principle of Maximum Conformality and Commensurate Scale Relations,''
Phys. Rev. D \textbf{89}, 014027 (2014).

\bibitem{Brodsky:1997jk}
 S.~J.~Brodsky and P.~Huet,
 ``Aspects of SU(N(c)) gauge theories in the limit of small number of colors'',
 Phys. Lett. B \textbf{417}, 145-153 (1998).

\bibitem{Gell-Mann:1954yli}
M.~Gell-Mann and F.~E.~Low,
``Quantum electrodynamics at small distances,''
Phys. Rev. \textbf{95}, 1300 (1954).

\bibitem{Shen:2017pdu}
J.~M.~Shen, X.~G.~Wu, B.~L.~Du and S.~J.~Brodsky,
``Novel All-Orders Single-Scale Approach to QCD Renormalization Scale-Setting,''
Phys. Rev. D \textbf{95}, 094006 (2017).

\bibitem{Yan:2022foz}
J.~Yan, Z.~F.~Wu, J.~M.~Shen and X.~G.~Wu,
``Precise perturbative predictions from fixed-order calculations,''
J. Phys. G \textbf{50}, 045001 (2023).

\bibitem{Wu:2018cmb}
X.~G.~Wu, J.~M.~Shen, B.~L.~Du and S.~J.~Brodsky,
``Novel demonstration of the renormalization group invariance of the fixed-order predictions using the principle of maximum conformality and the $C$-scheme coupling,''
Phys. Rev. D \textbf{97}, 094030 (2018).

\bibitem{Brodsky:2012ms}
S.~J.~Brodsky and X.~G.~Wu,
``Self-Consistency Requirements of the Renormalization Group for Setting the Renormalization Scale,''
Phys. Rev. D \textbf{86}, 054018 (2012).

\bibitem{Wu:2019mky}
X.~G.~Wu, J.~M.~Shen, B.~L.~Du, X.~D.~Huang, S.~Q.~Wang and S.~J.~Brodsky,
``The QCD renormalization group equation and the elimination of fixed-order scheme-and-scale ambiguities using the principle of maximum conformality,''
Prog. Part. Nucl. Phys. \textbf{108}, 103706 (2019).

\bibitem{Zheng:2013uja}
 X.~C.~Zheng, X.~G.~Wu, S.~Q.~Wang, J.~M.~Shen, and Q.~L.~Zhang,
 Reanalysis of the BFKL Pomeron at the next-to-leading logarithmic accuracy,
 J. High Energy Phys. {\bf 10}, 117 (2013).

\bibitem{Cacciari:2011ze}
M.~Cacciari and N.~Houdeau,
``Meaningful characterisation of perturbative theoretical uncertainties,''
JHEP \textbf{09}, 039 (2011).

\bibitem{Bagnaschi:2014wea}
E.~Bagnaschi, M.~Cacciari, A.~Guffanti and L.~Jenniches,
``An extensive survey of the estimation of uncertainties from missing higher orders in perturbative calculations,''
JHEP \textbf{02}, 133 (2015).

\bibitem{Bonvini:2020xeo}
M.~Bonvini,
``Probabilistic definition of the perturbative theoretical uncertainty from missing higher orders,''
Eur. Phys. J. C \textbf{80}, 989 (2020).

\bibitem{Duhr:2021mfd}
C.~Duhr, A.~Huss, A.~Mazeliauskas and R.~Szafron,
``An analysis of Bayesian estimates for missing higher orders in perturbative calculations,''
JHEP \textbf{09}, 122 (2021).

\bibitem{Du:2018dma}
 B.~L.~Du, X.~G.~Wu, J.~M.~Shen and S.~J.~Brodsky,
 ``Extending the Predictive Power of Perturbative QCD'',
 Eur. Phys. J. C \textbf{79}, 182 (2019).

\bibitem{Shen:2022nyr}
J.~M.~Shen, Z.~J.~Zhou, S.~Q.~Wang, J.~Yan, Z.~F.~Wu, X.~G.~Wu and S.~J.~Brodsky,
``Extending the predictive power of perturbative QCD using the principle of maximum conformality and the Bayesian analysis,''
Eur. Phys. J. C \textbf{83}, 326 (2023).

\bibitem{Shen:2023qgz}
J.~M.~Shen, B.~H.~Qin, J.~Yan, S.~Q.~Wang and X.~G.~Wu,
``Novel method to reliably determine the QCD coupling from R$_{uds}$ measurements and its effects to muon g \ensuremath{-} 2 and $ \alpha \left({M}_Z^2\right) $ within the tau-charm energy region,''
JHEP \textbf{07}, 109 (2023).

\bibitem{Luo:2023cpa}
Y.~F.~Luo, J.~Yan, Z.~F.~Wu and X.~G.~Wu,
``Approximate N$^{5}$LO Higgs Boson Decay Width $\Gamma(H\to\gamma\gamma)$,''
Symmetry \textbf{16}, 173 (2024).

\bibitem{Duhr:2019tlz}
C.~Duhr and F.~Dulat,
``PolyLogTools \textemdash{} polylogs for the masses,''
JHEP \textbf{08}, 135 (2019).

\bibitem{Bauer:2000cp}
C.~W.~Bauer, A.~Frink and R.~Kreckel,
``Introduction to the GiNaC framework for symbolic computation within the C++ programming language,''
J. Symb. Comput. \textbf{33}, 1 (2002).

\bibitem{Bi:2015wea}
H.~Y.~Bi, X.~G.~Wu, Y.~Ma, H.~H.~Ma, S.~J.~Brodsky and M.~Mojaza,
``Degeneracy Relations in QCD and the Equivalence of Two Systematic All-Orders Methods for Setting the Renormalization Scale,''
Phys. Lett. B \textbf{748}, 13 (2015).

\bibitem{Baikov:2016tgj}
P.~A.~Baikov, K.~G.~Chetyrkin and J.~H.~K\"uhn,
``Five-Loop Running of the QCD coupling constant,''
Phys. Rev. Lett. \textbf{118}, 082002 (2017).

\bibitem{Herzog:2017ohr}
F.~Herzog, B.~Ruijl, T.~Ueda, J.~A.~M.~Vermaseren and A.~Vogt,
``The five-loop beta function of Yang-Mills theory with fermions,''
JHEP \textbf{02}, 090 (2017).

\bibitem{Wu:2013ei}
X.~G.~Wu, S.~J.~Brodsky and M.~Mojaza,
``The Renormalization Scale-Setting Problem in QCD,''
Prog. Part. Nucl. Phys. \textbf{72}, 44 (2013).

\bibitem{DiGiustino:2023jiq}
L.~Di Giustino, S.~J.~Brodsky, P.~G.~Ratcliffe, X.~G.~Wu and S.~Q.~Wang,
``High precision tests of QCD without scale or scheme ambiguities,''
Prog. Part. Nucl. Phys. \textbf{135}, 104092 (2024).

\bibitem{Denner:1990cpz}
A.~Denner and T.~Sack,
``The W-boson width,''
Z. Phys. C \textbf{46}, 653 (1990).

\bibitem{Meng:2022htg}
R.~Q.~Meng, S.~Q.~Wang, T.~Sun, C.~Q.~Luo, J.~M.~Shen and X.~G.~Wu,
``QCD improved top-quark decay at next-to-next-to-leading order,''
Eur. Phys. J. C \textbf{83}, 59 (2023).

\end{thebibliography}
\end{document}